# Fast Fisher-Lee approach for conductance calculations on BTB-based molecular junctions: effects of isomerization and electrode coupling

Sylvain Pitié[1], Mahamadou Seydou[1*], Yannick. J. Dappe[2], Pascal Martin[1], François Maurel[1], Jean Christophe Lacroix[1]

[1] Université de Paris, ITODYS, CNRS, F-75006 Paris, France.

[2] Service de Physique de l'Etat Condensé (URA 2464 CNRS), Service de Physique et Chimie des Surfaces et des Interfaces, Laboratoire des Nano-Objets et Systèmes Complexes.

*mahamadou.seydou@u-paris.fr

# Abstract

In this work, we have implemented the Fisher-Lee formalism to couple non-equilibrium Green's functions with tight-binding Density Functional Theory (DFT) to tackle large molecular systems. This method is used to determine the decay constant of a set of oligomers based on seven different monomers taken from the literature in the non-resonant tunneling regime. Results show good agreement with experimental measurements.

The approach is then applied to explore the conformational pattern effect as well as the asymmetry and the strength of coupling with the electrode of X−(1-(2-bisthienyl) benzene)$_{n=1, 5}$)−Y (X, Y= gold (Au), titanium (Ti) and graphene (G)) junctions. The results indicate that conformational patterns have low impact on the conductance, since the delocalization of π-electrons exhibits similar behavior for all the conformations explored. The calculated attenuation factor is found to be comparable with the strength of contact coupling (Ti > Au ~ G).

Electronic analysis of metal-molecule interactions reveals the ionic nature of Ti-C bonds, through the emergence of a local dipole contributing to the work function variation of 0.35 eV. In addition, the Ti *d*-orbitals are found to be strongly coupled with the lowest unoccupied orbital (LUMO) of BTB, thus facilitating charge transfer from Ti to the molecule, at the origin of this strong interfacial dipole. However, the Au-C bond is found to be similar to the C-C bond, with pure covalent character. The results confirm the hole transport mechanism observed experimentally in the cases of Au-(BTB)n-Au and Au-(BTB)$_n$-Ti, and predict possible combined mechanism of both hole and electron transport in the case of Ti(BTB)$_n$-Ti.

**Keywords:** electron transport, NEGF, DFT, BTB, titanium, gold, graphene, attenuation factor, decay constant.

# 1- INTRODUCTION

The possibility offered by organic molecules to mimic electronic functions (resistance, diode, transistor, switch) is at the origin of the emergence of an active disciplinary field at the interface of Chemistry and Physics[1,2]. Organic and inorganic molecules can be modified relatively easily to constitute quantum objects that can deal with silicon technology[3,4]. They promise lower cost and probably lower electrical consumption. Molecular electronics aims at using molecular components to develop nano-electronic devices with tunable new function perspectives.

In this context many experimental setups have been developed with the aim of sandwiching selected molecules between electrodes and measuring their transport properties at the nanoscale[5–7]. Hence, while experimental setups seem to be well established, many points are still not understood and many obstacles have to be overcome before industrial production can be considered[4]. The problems in this area imply, for example, control of top molecular-electrode junction reliability of measurements, detailed understanding of transport mechanisms as a function of the junction length, etc.[4, 5] In this respect, intrinsic properties of molecular junctions like rectification, that is often considered to be related to the asymmetry of the junctions at the level of the electrodes[8], and quantum interference leading to a strong attenuation of the conductance in certain junctions[9], are still debated within the community.

To tackle these challenges, several theoretical and numerical approaches have been developed recently to accurately determine electron transport within metal-molecule-metal junctions using quantum calculations.[10] The common approach, based on non-equilibrium Green's functions (NEGF), has given results in very good agreement with experiment. In particular, NEGF combined with density functional theory (DFT) in a fully self-consistent process makes it possible to effectively treat the diffusion region coupled to the electrodes[11]. Simulation tools[6, 8, 9] are now reliable and accessible, and can provide useful information on electron transfer mechanisms in junctions of reasonable size. These methods are based on the Landauer-Büttiker approach[12] which expresses the conductance of a coherent system in terms of a quantum-mechanical scattering problem. Important improvements have been made lately in the way of describing the leads and in techniques for solving the scattering problem.[9, 10]

Recently, the coupling of experimental measurements with theoretical calculations has made it possible to interpret the transport mechanism of large ranges of conjugated molecular systems.[13] Among these systems, attention have been given to thiophene-based oligomers due to their favorable chemical stability, their synthetic versatility and their efficient charge transport along the highly conjugated π-orbitals of the oligomers.[14]

As such, several studies have been conducted on pure oligothiophenes and their side or terminal group functionalization.[15–18] Indeed, Tada's group performed several studies on the synthesis and transport properties of these oligomers and showed that the conductance varies exponentially with the oligomer length.[16,17] Conversely, other authors found a significant deviation in conductance with length due to conformational change for some oligomers.[18,19] Xiang *et al*.[15] used iodine to functionalize thiophene-based oligomers and found a non-exponential variation of the conductance with the molecular length, that they attributed to modification of the frontier orbital levels. Combining theoretical calculations and experiments, Leary *et al*. explored solvent effects on oligothiophenes and showed that the molecular conductance is over 2 orders of magnitude larger in the presence of a shell comprising 10 water molecules.[19] More recently, Li *et al*.[20] investigated charge transport through intermolecular and intramolecular paths in single-molecule and single-stacking thiophene junctions using the mechanically controlled break-junction (MCBJ) technique. They concluded that intermolecular charge transport offers the efficient and dominant path at the single-molecule scale. Even if experimental and theoretical developments progress, full understanding of the transport mechanism remains a challenge due to the time limits of calculations, the complexity of the oligomer conformations and their coupling to the electrodes.

In this work, we have implemented the well-known Fisher-Lee approach[21] in a DFT tight-binding code which is relatively fast compared to other codes. First, we use this approach to determine the attenuation factors of a range of molecular systems taken from the literature as benchmarks. In the non-resonant tunneling regime the conductance $G$ varies exponentially with the molecular length $L$ ($G = G_c e^{-\beta L}$), where $\beta$ is the attenuation factor, also called the decay constant, and $Gc$ is the contact conductance, equal to the inverse of the contact resistance $Rc$. The contact resistance $Rc$ is known to be primarily dependent on the surface structure and the specific interaction site between the molecule and the surface[22]. Experiments on molecular junctions are carried out under different conditions, and similarly, from a theoretical point of view, calculations are made on electrodes ranging from small clusters to the ideal periodic surface. Under these very different conditions, direct comparison between contact resistances is very delicate. Thus, we prefer to focus on the attenuation factor. The results obtained are compared with experiments and other theoretical studies as a benchmark of our implementation. Second, we explore the electron transport properties in linear polymeric oligo(1-(2-bisthienyl)benzene) (BTB) sandwiched between titanium, gold and graphene electrodes. These π-conjugated systems, using grafted oligothiophenes, were investigated experimentally in large-area junctions as well as at the single-molecule level. Their electronic transport properties in the non-resonant tunneling regime have been determined experimentally[23–25]. Finally, we used standard DFT calculations to interpret quantum transport results by evaluating the adsorption mechanism of anchoring groups on common stable surfaces (Au, Ti, graphene). This leads to a better understanding of the influence of molecule/electrode coupling on the attenuation factor.

## 2.1. QUANTUM TRANSPORT CALCULATIONS

Calculations have been performed within Density Functional Theory (DFT) formalism as implemented in the Fireball code[26]. The Fireball code is a very efficient tool using an optimized localized orbital basis set[27], and the local density approximation (LDA) for the exchange and correlation energy through the McWeda formalism[28]. It offers a good compromise between computational time and accuracy, allowing to study large hybrid organometallic systems. A Keldysh-Green approach was previously implemented in the Fireball code[29] and used mainly for STM image calculations, but also to model molecular junctions[30]. In this approach, the system is decomposed into two parts, namely the tip and the substrate[29]. This approach is limited by the choice of the frontier between the two parts, usually the bottleneck for the current flowing in the system. A possible solution consists in using the Fisher-Lee approach[21] where the molecular part and the two electrodes are clearly separated.

## 2.2. IMPLEMENTATION OF FISHER-LEE APPROACH IN THE FIREBALL CODE

The implementation of Fisher-Lee formalism consists in decomposing the system into three parts (see scheme 1), namely the left and right electrodes and the central part, *i.e.* the molecule. The Green's functions of Right and Left parts are defined by:

$$G(E)_{R,L} = [(E - i\eta)I - H_{R,L}]^{-1} \qquad (1)$$

and the Green's function of the central part can be written as:

$$G_C(E) = [(E - i\eta)I - H_C - \Sigma_L(E) - \Sigma_R(E)]^{-1} \quad (2)$$

where $\Sigma_{L,R}(E)$ are the self-energies of the electrodes and η is the imaginary part of the Green's function which is representative of the electronic level width or, in other words, the coupling between these levels. This parameter is used to modify the level widths of a specific subsystem, and in particular allows to distinguish the electron reservoir and the molecular channel. Hence, on the one hand, the electron reservoir levels are enlarged, to take into account the bulk character, and on the other hand, the molecular electronic level widths are reduced to represent the discrete molecular levels.

Consequently, the transmittance $T(E)$ is related to the Green's functions through:

$$T(E) = Tr[\Gamma_L(E)G_C(E)\Gamma_R(E)G_C^\dagger(E)] \quad (3)$$

where the $\Gamma_{R,L}$ matrices correspond to the coupling matrices between the electrodes and the central part. The transmittance spectrum can therefore be plotted for each optimized system at zero bias.

In practice, we first optimized the geometries of a family of seven molecular junctions gathered from the literature. Once the equilibrium position was found, we determined the corresponding electronic structure by density of states (DOS) calculations. From these DOS and the hopping integrals of the system, we performed electron transport calculations. We choose a value of η of 0.5 eV for the electrode part and $10^{-8}$ eV for the molecular part.

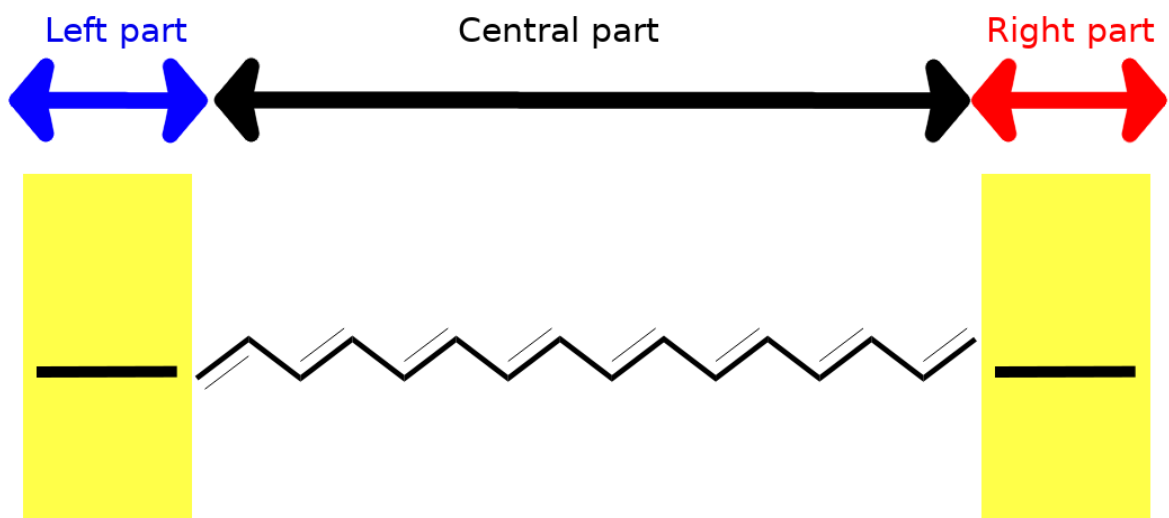

Scheme 1: Decomposition of the junction into three parts.

## 2.3 PURE DFT/PBE CALCULATIONS

To analyze the nature of anchoring group bonding, we performed DFT calculations under the generalized gradient approximation (GGA) with the Perdew-Burke-Ernzerhof (PBE) functional[31] as implemented in the Vienna Ab-initio Simulation Package (VASP 5.4.1)[32] in order to model the binding of benzene and thiophene groups on Au(111), Ti(001) and graphene-based electrodes. To this end, electron-ion interactions were described by the projector augmented wave (PAW)[33] method, representing the valence electrons, as provided in the code libraries. The convergence of the plane-wave expansion was obtained using a cut-off energy of 500 eV. To adequately describe the effects of van der Waals interactions, all the computations reported in this work were performed using the dispersion-included DFT-D3 method.[34] For geometry optimization, sampling in the *Brillouin* zone was achieved on a (3 × 3 × 1) grid of k-points.

In the present study, slabs representing the Au (111) and Ti (001) surfaces were cut out of the optimized bulk face-centered cubic cell of gold and a hexagonal titanium structure, respectively, using Modelview software.[35] For Au, the face-centered cubic (fcc) bulk optimized cell parameter at the PBE level was 4.08 Å, in good agreement with experiment (4.08 Å).[36] For Ti, the hexagonal bulk optimized parameter was *a = b = 2.95 Å* and *c = 4.68 Å* in agreement with crystallographic results.[37]

Both surfaces are modeled as a slab, where a unit cell is periodically reproduced in two dimensions *(x,y)*, with a vacuum space in the *z-axis* direction. This vacuum space height is set to 40 Å, enough to enable molecule biding and disable its interaction with the consecutive repetition of the system. In the case of Au and Ti, the slab model consists of four layers, where the two bottom ones are frozen in the optimized bulk positions and the two upper layers are relaxed. A super-cell representing a (3 x 3) super-cell is built from the optimized Au(111) and Ti (001) and (5 × 5) unit cells for graphene. This super-cell size allows to explore more finely the adsorption of ternary (hollow), binary (bridge) and top sites (scheme 2).

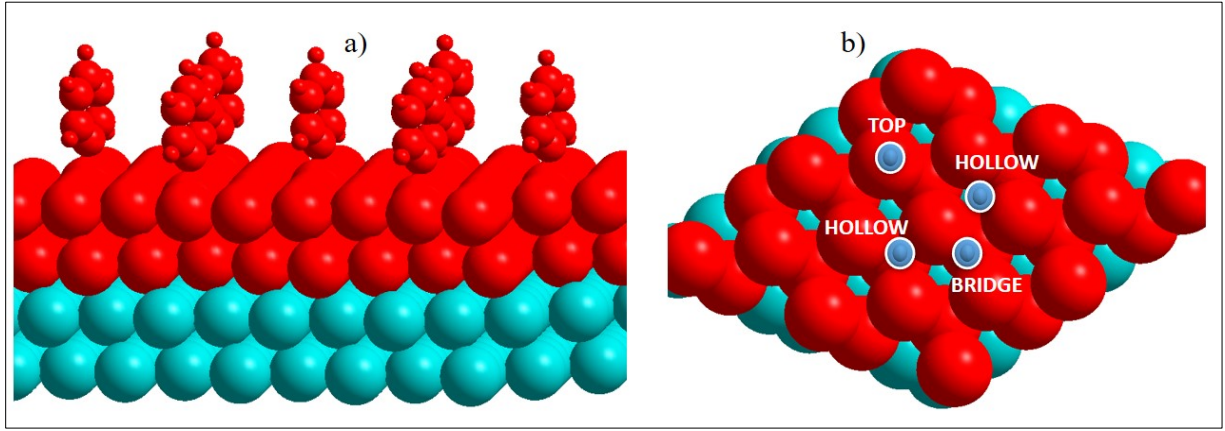

Scheme 2: (a) molecule adsorbed on Au (111) super-cell showing relaxed atoms (red) and fixed atom (blue) during geometry optimization. (b) Top view of the (3 × 3) super-cell indicating possible adsorption sites.

The binding energy is computed as the difference between the energies of optimized metal-molecule and the isolated bare metal and molecules optimized separately.

$$\Delta E_{bind} = E(Mm) - E(M) - E(m)$$

where *E(Mm)* is the optimized energy of the metal-molecule interface, *E(m)* is the optimized energy of the isolated molecule and *E(M)* is the energy of the bare surface.

In order to analyze bond formation, we have determined from the optimized structures the charges densities of complexes and of separate molecules and surfaces ($\Delta\rho(r) = \Delta\rho_{Mm}(r) - \Delta\rho_M(r) - \Delta\rho(r)_m$) with high precision, by sampling the Brillouin zone on a grid of (9 x 9 x 1)k-points. In addition, we have calculated the Bader charges[38] on this fine grid of charge densities. Finally, we computed the work function and its variation, to highlight the formation of an interface dipole.

## 3. RESULTS AND DISCUSSION

### 3.1 FISHER-LEE APPROACH BENCHMARK

The new methodology implemented is applied to compute the decay constant and contact resistance of molecular junctions constructed from a set of seven monomers shown in Figure 1a sandwiched between gold electrodes. These junctions are chosen according to the availability of experimental measurements of the decay factor. The structures as well the anchoring groups are different. Using DFT formalism as implemented in the Fireball code[26], we built the structures of the oligomers and fully optimized their geometries. From these optimized geometries, the transmittance spectra are computed using (NEGF + DFT) within the Fisher-Lee approach. In the non-resonant tunneling regime, since the transmittance $T(E)$ varies exponentially with the length $(L)$ of the oligomer chain, the logarithm of the transmittance $Log(T(E))$ varies linearly with L ($Log(T(E)) = -\beta L + Log(1/R_c)$). In Figure 1b, $Log(T(E))$ plotted as a function of $L$ shows a linear decrease, as expected for all systems. The calculated decay constant and contact resistance are deduced from the linear fit of this line. The calculated values of $R_c$ are compared with the experimental and theoretical ones found in the literature in Table S1. As we said earlier in the Introduction, contact resistance $R_c$ is not discussed here.

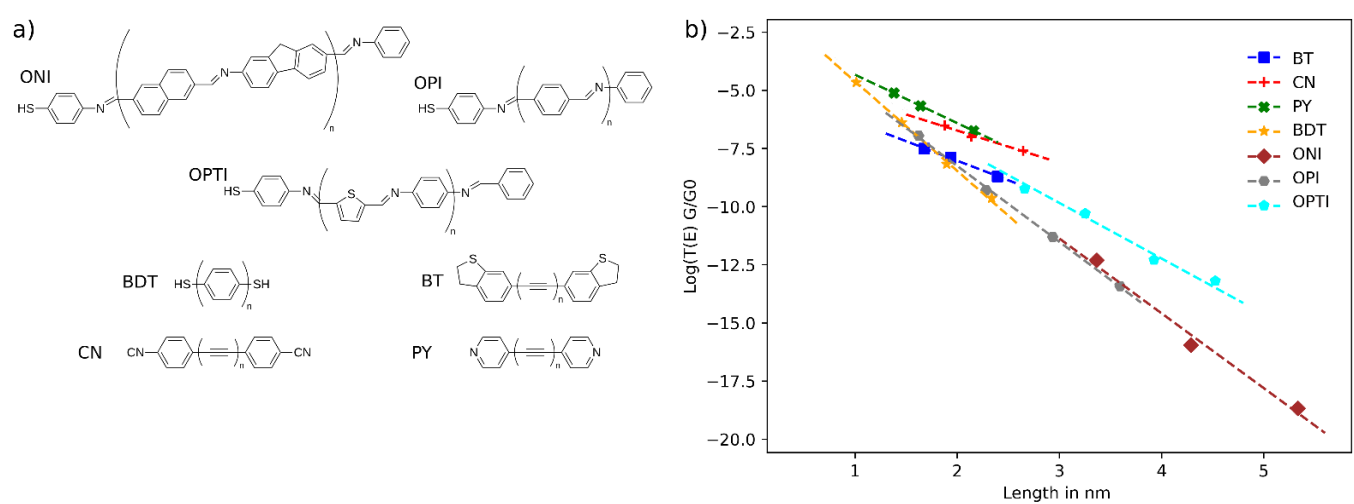

*Figure 1: a) Representation of the molecular systems considered for the benchmark: ONI = oligonaphthalenefluoreneimine, OPI = oligophenyleneimine, OPTI = oligophenylene-thiopheneimine, BDT = oligophenylene, BT = dihydrobenzothiopheneoligoyne, CN = cyanoligoyne, PY = pyridyloligoyne. b) Plot of the corresponding transmittance at the Fermi level, as a function of the molecular length.*

The systems investigated can be divided in two groups: On the one hand, ONI, OPI and OPTI and BDT were investigated using the conductive-probe atomic force microscopy (CP-AFM) technic. They are grafted onto the electrode by a thiol group forming the well-known Au-S bond. Theoretical calculations are carried out on OPI and BTD within the framework of Landauer-Buttiker formalism. On the other hand, diaryloligoynes functionalized by BT, CN and PY as anchoring groups were studied experimentally by the scanning tunneling microscopy break-junction (STMBJ) method.

Table 1 summarizes the values of the decay constant $\beta$ (in nm$^{-1}$) compared to experimental and other theoretical studies found in the literature. Our calculated values are in good agreement with experiment for all the systems considered, with a deviation less than 1.0 nm$^{-1}$. Overall, theoretical calculations underestimate the *β* value compared to experimental measurements. This has been observed previously and attributed to the well-known trend of DFT to underestimate the band gap, leading to an underestimation of the decay constant.[39] In addition, the discrepancies between our values and experimental ones seem to be related to the structure of the anchored group. In the first series, the maximum deviation (0.8 nm$^{-1}$) is found for OPTI[40] for which the anchor group with the top electrode is a phenyl ring. In the second series, the highest deviations are obtained for the BT and PY cyclic anchoring groups with 1.2 and 1.0 nm$^{-1}$, respectively. These cyclic groups make the description of the molecule-electrode coupling less consistent. The differences with the other theoretical results originate also from the nature of the electrodes that are modeled by clusters in this work or periodic ideal surfaces in other cases.[39]

| Systems | $\beta_{exp}$ (nm$^{-1}$) | $\beta_{theo}$ (nm$^{-1}$) | $\Delta\beta_{theo-exp}$ (nm$^{-1}$) | $\beta_{this\ work}$ (nm$^{-1}$) | $\Delta\beta_{this\ work-exp}$ (nm$^{-1}$) |
|---|---|---|---|---|---|
| ONI | 2.5[41] | / | / | 3.2 | 0.7 |
| OPI | 3.0[41] | 2.5[42] | 0.5 | 3.1 | 0.1 |
| OPTI | 3.4[40] | / | / | 2.6 | 0.8 |
| BDT | 4.2[43] | 1.3[44] | 2.9 | 3.9 | 0.3 |
| BT | 2.9[39] | 1.7[39] | 1.2 | **1.7** | **1.2** |
| CN | 1.7[39] | 0.4[39] | 1.3 | 1.4 | 0.3 |
| PY | 3.1[39] | 2.2[39] | 0.9 | 2.1 | 1.0 |

Table 1: Summary: decay constant *β* compared to experimental data and other theoretical calculations.

## 3.1. OLIGOMERS CONNECTED TO GOLD ELECTRODES: ISOMERIZATION EFFECT, ENERGETICS AND ELECTRONIC TRANSMITTANCE

The influence of geometrical modifications on the electronic transport properties was investigated for BTB dimer conformers. This latter is connected to one electrode through the phenyl group in the *para* position. It is connected to the other electrode through the thiophene carbon in the α position. The conformation search consists in a simple 180° rotation of one thiophene ring with respect to the other. This leads to the formation of *cis* and *trans* conformations. Since the dimer considered here is made up of four thiophenes, we consider three conformations corresponding to *trans-trans-trans*, *cis-cis-cis* and *cis-trans-cis*, presented in Figure 2c.

After geometry optimization, the results show that all conformers are planar. The *trans-trans-trans* form is the most stable. The energy differences between the conformers are small and less than 150 meV. The most stable conformer is (*trans-trans-trans*) due to the minimization of steric hindrance of the hydrogen atoms on two adjacent molecules. The rotation barrier computed previously by Lin *et al.*[45] was about 108 meV. This low value suggests that the formation of these conformers as possible at room temperature and under the experimental conditions.

The transmittance spectra of the three conformers of the *para* isomer, named (*cis-trans-cis*), (*trans-trans-trans*), and (*cis-cis-cis*), are presented in Figure 2b. The difference in transmittance is minor, which means consequently that the transmittance depends mainly on the delocalization of the π-electrons, which is similar in all the conformers. On the contrary, the 90° rotation, involving the decoupling of π-orbitals, significantly affects the transmittance[46]. It was shown recently[17] for oligothiophenes that the *cis* conformer has a higher conductance than the *trans* one. This was explained by the highest occupied molecular orbital (HOMO) of the *cis* which is found to be closer to the Fermi level of the electrodes. In this case, the BTB (*cis-cis-cis*) conformer has also a higher conductance than the *trans*. unfortunately, the frontier orbitals are in the same position and cannot explain the difference. We suspect that the distance between the electrodes, imposed by the molecular conformation, is the reason for this variation. As can be seen in Figure 2c, the whole *cis* conformation induces a deformation which reduces the tunneling distance between the electron reservoirs.

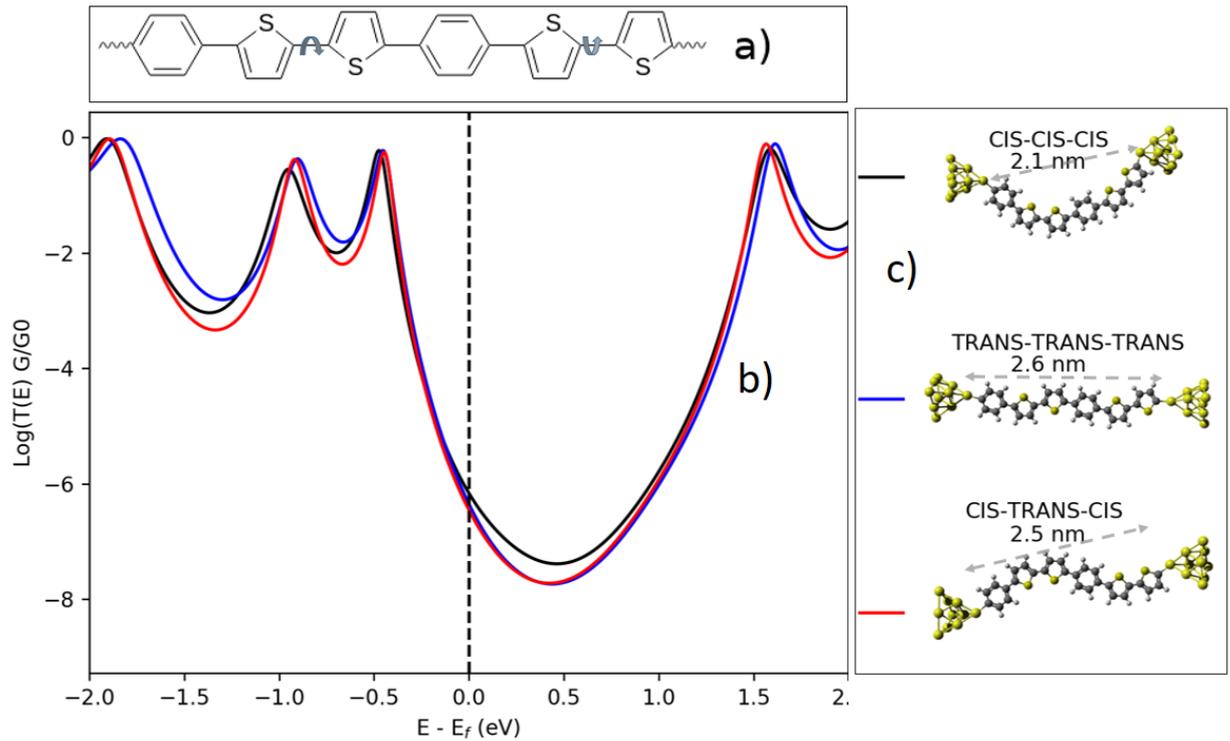

Figure 2: (a) Structure of BTB oligomers indicating isomer rotation angle. (b) Optimized geometries of (BTB)$_2$ isomers. (c) Transmission spectra of BTB isomers.

## 3.2 CONSTANT DECAY CALCULATION FOR ASYMMETRIC JUNCTIONS: STRENGTH OF COUPLING EFFECT

The decay constant or attenuation factor characterizes the conduction capacity of a molecular chain. This quantity depends both on the nature of the molecular system and on its connection to the electrodes. While it has been shown that the value varies for conjugated molecules between 2 and 3 nm$^{-1}$, recently, it has been proved that the coupling strength to the electrodes allows to modulate its value[47].

Here, the attenuation coefficient is computed through transport property determination for BTB-based oligomers sandwiched between gold (Au) and different electrodes. In particular, the phenyl group of BTB is anchored to gold and the thiophene terminal group is attached to gold, titanium (Ti), and graphene (G) electrodes. The results are compared to experiments carried out on Au-(BTB)$_n$-Ti[48] and G-(BTB)$_n$-G[24].

Figure 3 shows the variations of transmittance on a logarithmic scale (log ($T(E)$)) for different cases of (BTB)$_2$ junctions when the right electrode is modified. The transmittances at the Fermi level are 2 10$^{-2}$, 2 10$^{-3}$, 2 10$^{-4}$ and 5 10$^{-7}$ G$_0$ for Ti, Au, G$_{cov}$ and G$_{vdW}$, respectively. From these results, we can deduce that the coupling to the titanium electrode leads to the highest probability of electronic transmission. It is followed by gold and graphene covalently bonded to BTB. The lowest transmittance is observed for the case where the BTB is weakly bonded to graphene through van der Waals interactions. In this last case, an additional tunnel barrier is added between the electrodes due to the lack of a support channel between the anchoring unit and the graphene electrode.

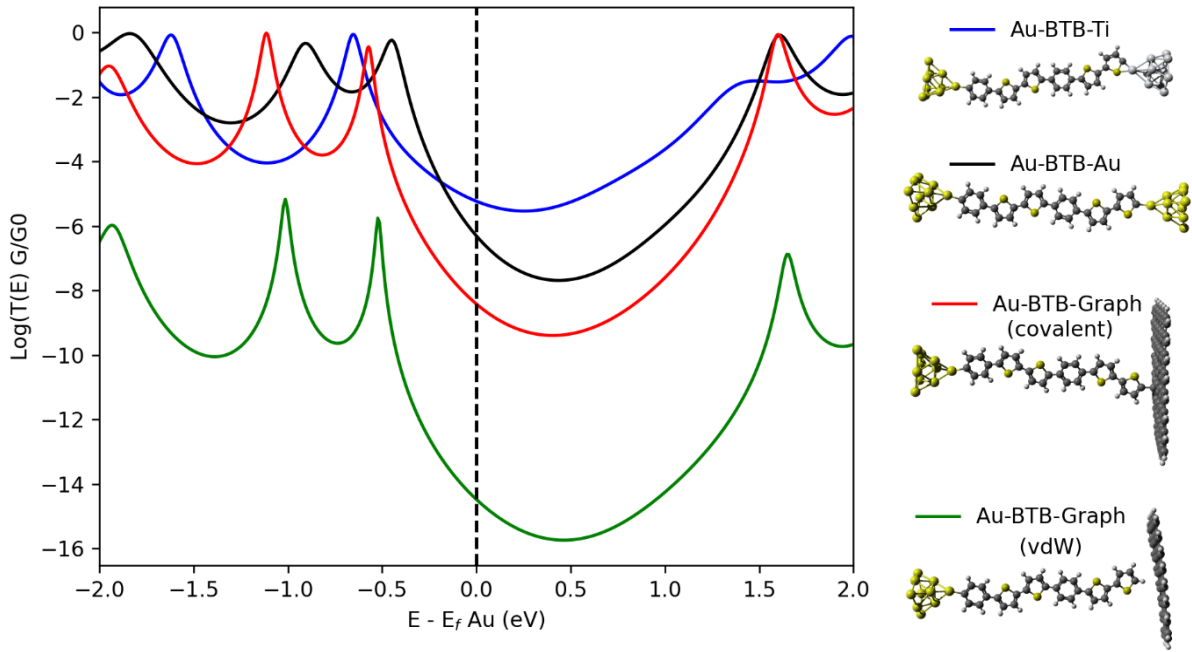

Figure 3: Variation of *T(E)* of *(BTB)$_2$* for different right electrodes.

In Figure 4a, the transmittance value at Fermi level ($T(E_f)$) is plotted as a function of the molecular length, varying from 1 to 5 BTB units. As a result, all these asymmetric junctions provide a linear decrease (on a logarithmic scale) of conductance in the non-resonant tunneling transport mechanism. The ordinates at the origin represent the contact resistances: 46.1, 21.9, 258.1 and 3.2 $G_0$ for Au, Ti, $G_{cov}$ and $G_{vdW}$, respectively.

The slopes of the curves represent the attenuation factors $\beta$ : 1.8, 2.0, 2.1 and 2.1 nm$^{-1}$ for Ti, Au, $G_{cov}$ and $G_{vdW}$, respectively. The calculated value for Ti is in a good agreement with experimental measurements[48] (1.8 nm$^{-1}$). McCreery *et al.*[24] found a $\beta$ value around 2.9 nm$^{-1}$ for G-(BTB)$_n$-G, which is similar to Au-(BTB)$_n$-G$_{cov}$ (2.1 nm$^{-1}$) and Au-(BTB)$_n$-Au (2.0 nm$^{-1}$) presented on Figure S2. The decay constant depends on both the coupling between electrodes and anchoring groups (type and strength) and the frontier orbital (HOMO, LUMO) positions.

In Figure 4b, the variation of the binding energy ($\Delta E_{bind}$) with the decay constant shows that the lower $\Delta E_{bind}$, the lower is $\beta$, in agreement with the conclusions of a recent general review on molecular junctions, which claims that $\beta$ depends mainly on the strongly coupled electrode.[47] Indeed, for stronger coupling, one can expect a better conduction, in agreement with the transmittance value at the Fermi level presented above. The lowest constant is found for the Ti electrode with a binding energy of -4.885 eV and the highest for graphene non-covalently bonded to the thiophene anchoring group, with a binding energy of -0.522 eV. While there is no result in the literature on molecular adsorption on Ti surfaces, the binding energy obtained for benzene on $G_{cov}$ is -1.617 eV, in agreement with the literature.[49] On Au (111) surface, it is -2.674 eV, in good agreement with previous studies[50,51].

Figure 4c presents the position of the lowest unoccupied orbital (LUMO) of $(BTB)_2$ which is located at 1.46, 1.32, 1.53 and 1.52 eV above Fermi energy level ($E_F$) for Au, Ti, $G_{cov}$ and $G_{vdW}$, respectively. The highest occupied orbital (HOMO) is located at 0.51, 0.67, 0.52 and 0.53 eV below $E_f$ for Au, Ti, $G_{cov}$ and $G_{vdW}$, respectively. The LUMO of Ti is close to the Fermi level, making it the best conduction channel in the case of Au-$(BTB)_2$-Ti. This means that the coupling to the electrode has an influence on the acceptor/donor character in the molecular junction.

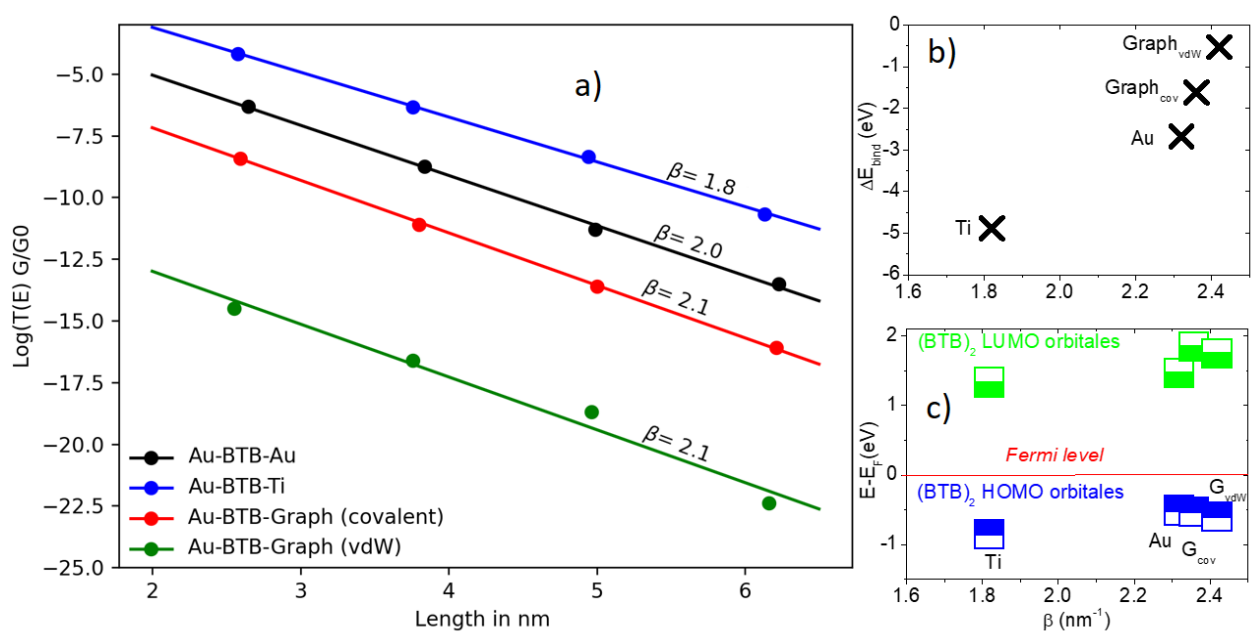

Figure 4: *Log (T(E$_F$))* as a function of the molecular junction length (a); variation of ΔE$_{bind}$ relative to attenuation factor β (b); HOMO and LUMO orbitals in function of β.

From these results, one can observe that the relation $\Delta E_{bind}$ versus $\beta$ appears to be linear for Au and graphene junctions. However, titanium shows different behavior with respect to its strong interaction with the molecule and coupling of Fermi level with the frontier orbitals of BTB (Figure 5). The total density of states (DOS) projected onto the d-valence orbitals of Ti ($d_{Ti}$) and Au ($d_{Au}$) (Figure S3) shows for $d_{Ti}$ a broad band that covers both HOMO and LUMO around the Fermi level while $d_{Au}$ interact merely with the HOMO below the Fermi level. We changed the positions of the Au and Ti electrodes and studied a symmetrical junction with only Ti electrodes. The corresponding results (Figures S1 and S2) support the hypothesis of a stronger interaction of titanium with BTB resulting in a displacement of LUMO toward the Fermi level of titanium. These observations are in agreement with the difference in electronegativity between Ti and carbon, suggesting an electron transfer from titanium to the carbon atom of BTB. This charge transfer process is investigated in the next section.

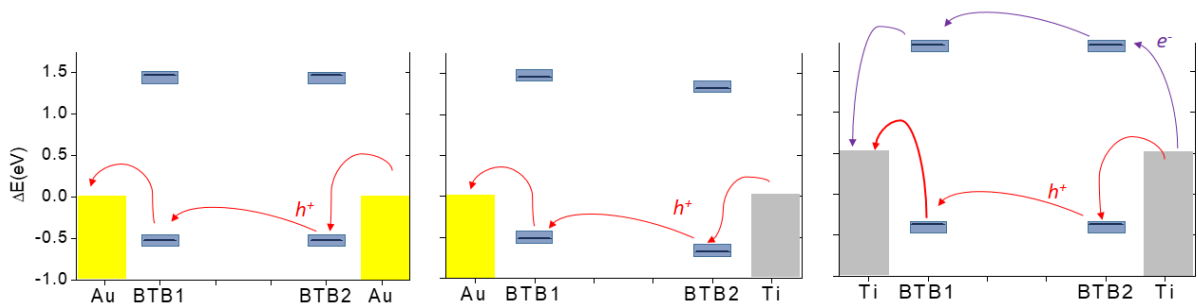

Figure 5: Position of (BTB)$_2$ HOMO and LUMO orbitals relative to metal Fermi levels for Au-(BTB)$_2$-Au (a), Au-(BTB)$_2$-Ti (b) and Ti-(BTB)$_2$-Ti (c).

### 3.3. ANALYSIS OF THE NATURE OF AU/TI/G–C BONDING

The similar behavior observed in the case of Au and G can be explained by the nature of the chemical interaction between the electrode and the anchoring group. Indeed, by calculating the Bader atomic charges and the differences in charge densities (Table S2), we have shown that the Ti-thiophene bond is covalent with a significant ionic character. Indeed, calculations show a charge transfer from Ti to the molecule (0.5 e-) in agreement with the shift of LUMO position toward Fermi level energy.

Conversely, for gold or graphene charge transfer with the anchoring group is weak. These results are in line with the difference in electronegativities. The electronegativity difference between titanium and carbon is 1.05 while it is 0.01 between gold and carbon.

| Metal-Molecule | Binding energy (eV) | C-M bond length (Å) | Sum of covalent radii (Å) | Ionic character (%) | $\Phi_{surface}$ (eV) | $\Phi_{sam}$ (eV) | $\Delta\Phi$ (eV) | $\Delta V_m$ | BD |
|---|---|---|---|---|---|---|---|---|---|
| 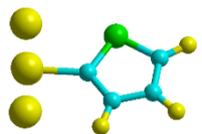 Au--Thiophene | -3.183 | 2.041 | 1.99 | 2.5 | 5.19 | 5.11 | -0.08 | -0.01 | -0.07 |
| 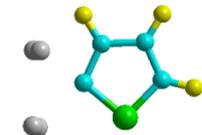 Ti--Thiophene | -4.885 | 2.37 | 2.11 | 11.1 | 4.37 | 4.09 | -0.28 | -0.01 | -0.27 |
| G--Thiophene 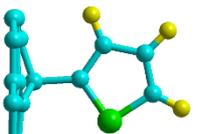 | -1.623 | 1.56 | 1.50 | 3.2 | 4.22 | 4.25 | 0.03 | -0.01 | 0.04 |

Table 2: Adsorption energies, thiophene-surface distance, bare metal ($\Phi_{surface}$) and complex ($\Phi_{sam}$) work functions, $\Delta V_m$ and BDE for Au, Ti and G electrodes.

These results prove that the Ti-C bond presents significant iono-covalent character, while the ionic character of the Au-C bond is weak. We have calculated the variation of the metal work function due to its interaction with the anchoring group. It was shown and largely discussed in the literature[52] that this variation can be decomposed into two terms[53]: $\Delta\Phi = \Delta V_m + BDE$, where $\Delta V_m$ is the variation of the potential between the anchoring group and the top of the molecule, and BDE is the variation of potential due to charge redistribution between the anchoring group and the metal surface. The two quantities were computed at the DFT level and the results are reported in Table 2. As can be seen, the BDE value is be close to zero for Au and graphene electrodes, in agreement with the lack of a local dipole at the interface, because of the similar electronegativities of the binding atoms. However, the BDE value is -0.35 eV for the Ti electrode, consistent with the observed charge transfer between Ti and C atoms. This corresponds to a strong surface dipole due the charge redistribution at metal molecule interface.

This redistribution coupled to the shift of LUMO may be at the origin of the asymmetry of the electronic current observed previously for Au-(BTB)$_n$-Ti junction[52,54] was demonstrated experimentally. Our calculations provide qualitative and quantitative informations supporting the assumptions made about the coupling strength and the positions of the frontier orbitals relative to the Fermi level. Indeed, strong adsorption energies are found for anchoring groups on Au and Ti surfaces. The Figure 5a presents the position of HOMO and LUMO of BTB relative to Au Fermi level showing a gap of 0.5 and 1.46 eV, respectively. This indicates that the preferential conduction channel would be the HOMO highlighting a mechanism of conduction by hole. The Figure 5b exhibits same information for the asymmetric Au-(BTB)$_2$-Ti junction. In this case, the LUMO moves toward Fermi level ($E_F$) of titanium, while the HOMO shifts away from. The gap becomes 1.33 eV between LUMO and $E_F$, and 0.67 eV between HOMO and $E_F$. The preferential conduction channel remains the HOMO predicting a mechanism of conduction by hole. In the case of Ti-(BTB)$_2$-Ti junction, as it can be seen on Figure 5c the gaps LUMO-$E_F$ and HOMO-$E_F$ are 1.32 and 0.87 eV, respectively. The shift of the LUMO toward $E_F$ is in agreement with the observed charge transfer from Ti to BTB molecule anchoring group. In this latter case, a conduction mechanism by electron becomes probable even if it may be minority.

**CONCLUSION**

In this work, we have implemented a fast powerful methodology, to determine the transport properties of large molecular junctions, within the DFT formalism coupled to NEGF using the Fisher-Lee formalism. A comparison with both experiment and other theoretical methodologies shows that this approach is very efficient.

We use this methodology to explore the geometries and electron transport properties of BTB oligomers sandwiched between gold, titanium and graphene electrodes. The results show that the rotation between monomers leading to *cis* and *trans* configurations exhibit similar transport properties.

The calculated values of the attenuation factor *β* are in a good agreement with experimental results. To the best of our knowledge, this is the first theoretical study on asymmetric X-(BTB)$_n$-Y (X. Y= Au, Ti, G$_{cov}$ and G$_{vdW}$) molecular junctions. We calculated the attenuation factor for the Au-(BTB)$_n$-Ti and G-(BTB)$_n$-G junction and found values of 1.8 and 2.1 nm$^{-1}$, respectively, very close to the experimental ones. The computed structural and electronic details confirm the experimental observations which announced similar transport properties for the junctions of Au-(BTB)$_n$-G and G-(BTB)$_n$-G. Our results indicate that the variation of $\beta$ depends slightly on the asymmetry of the electrodes and strongly on the nature and strength of the metal-molecule interaction.

The variation of $\beta$ roughly follows the calculated binding energies of molecules on metals. Analysis of the nature of the molecular bond shows that gold and graphene form the same type of quasi-covalent bond, in agreement with their respective electronegativities. For titanium, we have found a mechanism for charge transfer from the surface to the molecule in agreement with a very high interface dipole and the shift of the LUMO toward Fermi level. This sheds light on the asymmetric dependence of conductance on potential $G = f(V)$, at the origin of current. These results confirm and rationalize the hole transport mechanism proposed to interpret the current rectification observed experimentally for the Au-(BTB)$_n$-Ti junction and predicts a possible combined electron and hole transport mechanism for Ti-(BTB)$_n$-Ti.

## ASSOCIATED CONTENT
### SUPPORTING INFORMATION

Variation of $T(E)$ of (BTB)$_2$ for X-(BTB)$_2$-Y (X, Y= Au, Ti, G$_{cov}$ and G$_{vdW}$); Variation of $T(E_f)$ as a function of molecular length for different electrodes; Partial density of states of Au Ti and BTB; Comparison of experimental and calculated contact resistances; Bader charges analysis on binding sites and the difference in charge densities indicating the regions of charge accumulation and depletion.

## AUTHOR INFORMATION
### CORRESPONDING AUTHORS
*Mahamadou Seydou : **mahamadou.seydou@univ-paris-diderot.fr**

## ACKNOWLEDGEMENTS

Quantum chemical calculations were performed using HPC resources from GENCI-[CCRT/CINES/IDRIS] (Grant 2020[A0080807006].

ANR (Agence Nationale de la Recherche) and CGI (Commissariat à l'Investissement d'Avenir) are gratefully acknowledged for their financial support of this work through Labex SEAM (Science and Engineering for Advanced Materials and devices): ANR-10-LABX-096 and ANR-18-IDEX-0001.

**NOTES**

The authors declare no competing financial interest.

**TOC**

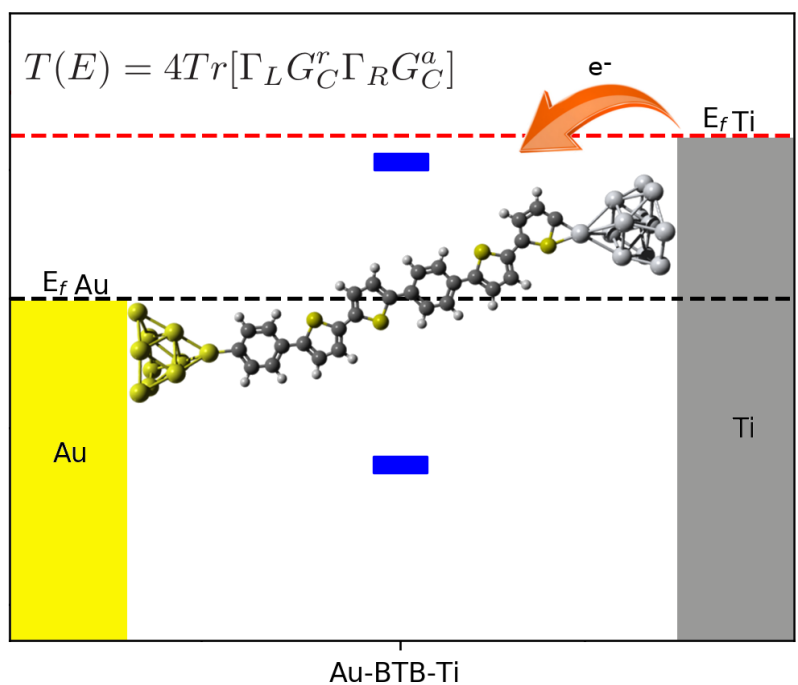

Au-BTB-Ti